\documentclass[prb, twocolumn, floatfix,superscriptaddress]{revtex4-2}
\usepackage{graphicx}
\usepackage{epstopdf}
\usepackage{amsmath, amsfonts, amssymb, bm,color}

\def\ifa{Institute of Applied Physics, Moldova State University,
Academiei 5, MD-2028 Chi\c{s}in\u{a}u, Moldova}

\def\ifin{National Institute of Physics and Nuclear Engineering, Reactorului 30, RO-077125, M\u{a}gurele-Bucharest, Romania}

\begin{document}
\title{Two-quanta processes in coupled double-quantum-dot cavity systems}
\author{Tatiana Mihaescu }
\email{mihaescu.tatiana@theory.nipne.ro}
\affiliation{\ifin}

\author{Aurelian Isar }
\email{isar@theory.nipne.ro}
\affiliation{\ifin}

\author{Mihai A. Macovei }
\email{mihai.macovei@ifa.usm.md}
\affiliation{\ifa}

\date{\today}
\begin{abstract}
The quantum dynamics of a compound sample consisting from a semiconductor double quantum dot (DQD) system non-linearly coupled with 
a leaking single-mode microresonator is theoretically investigated. The focus is on the resonance condition when the transition frequency 
of the DQD equals to the doubled resonator’s frequency, respectively, and the resulting interplay among the involved phonon or photon 
channels. As a result, the steady-state quantum dynamics of this complex non-linear system exhibits a variety of possible effects that 
have been demonstrated here. Particularly, we have found the relationship among the electrical current through the double quantum dot 
and the microwave field inside the resonator that is nonlinearly coupled to it, with a corresponding emphasising on their critical behaviours. 
Additionally, the quantum correlations of the photon flux generated into the resonator mode vary from super-Poissonian to Poissonian 
photon statistics, leading to single-qubit lasing phenomena at microwave frequencies.
\end{abstract}
\maketitle

\section{Introduction}
Interaction between light and matter have been long envisaged as a promising platform for novel applications \cite{expp1,expp2,theor}. Especially 
with the advancement of quantum information sciences, or generally quantum technologies, most experimental and theoretical efforts have been 
directed towards achieving real progresses in quantum optics and photonics. Optical cavities coupling natural atoms and photons or photonic on-chip 
devices with artificial atoms, or quantum dots, acting as emitters and detectors of light, are already manufactured, for instance. In this respect, 
investigations regarding interactions between qubits, defined as double quantum dots, and microwave photons are currently attracting growing 
attention. Particularly, strong-coupling regimes which manifest higher degrees of tunabilities of semiconducting DQDs are demonstrated in 
\cite{stokk,scar,khan,stcp1,stcp2}. The DQDs are typically coupled to normal or superconducting electronic reservoirs such that electronic 
transport may be used to characterize or modify the properties of the microwave cavity \cite{liuI,stock,haldar, hhhaldar, hhaldar}. Furthermore, 
the itinerant microwaves are efficiently detected in this configuration, by converting the photons into the charge current through DQDs 
\cite{khan, stanis}. Likewise, the photocurrent in the DQD can be probed by the microwave cavity, where the qubit's characteristics are 
recovered in the photon statistics \cite{havir, bur, burk, kul}.

Actually, this resulted in numerous theoretical investigations exploring the setup composed of DQDs embedded in various environments or 
coupled to leaking microwave resonators \cite{rev_mod,rev_art,elph1,elph2,agr2,chen,ghirri,nian,tarak}. For instance, in the cavity driven 
configuration the photon-assisted quantum transport through the DQD predicts high efficiency of microwave photon detection by studying 
the fluctuations of the photocurrent through the DQD in the mean-field approach neglecting dot-resonator correlations \cite{xu, wong, zenelaj}. 
More complex setups may include a Kerr non-linearity or consider cavity driving signals taken as single microwave pulses \cite{wang, nian, zenelaj2}. 
Also, the quantum spin transport is covered in \cite{gu} by considering superconducting leads. 

On the other hand, the manipulation of the photon emission spectrum and photon statistics through an electrical mean is also very desirable 
application for quantum communication and sensing representing a promising environment for photon statistics engineering in circuit-quantum 
electrodynamics \cite{gu,nian2, nian3, gul3}. Due to feasibilities of achieving strong coupling between electronic properties in semiconductor 
DQDs and the microwaves confined in micro-resonators, the maser effects may occur, which is an analog to lasing phenomenon but in 
gigahertz regimes \cite{mas1,gul,hartk,sql2,sql3,lamb,rastM,jin,chlo}. Moreover, a DQD driven out of equilibrium by a bias voltage acts 
as a gain medium for the microwave cavity and, within a narrow resonance window, a lasing state in the resonator is created, which 
highly correlates with the transport properties \cite{lamb,rastM,agr1,taba, jin, jin2}. Yet, the overall gain in this device is dominated 
by the phonon-assisted gain since a large contribution comes from the simultaneous emission of a photon and a phonon 
\cite{gul,hartk,kar}. The electron-phonon coupling leading to the occurrence of inelastic processes in the DQD coherently coupled 
with a microwave resonator is comprehensively covered in \cite{gul2}, while a driven DQD damped by the interaction with an 
unstructured phonon bath is presented in Ref.~\cite{st}, respectively. Interestingly, two qubit quantum correlations are obtained 
by entangling two DQDs by means of a transmission line microwave resonator \cite{contr} or a pair of laser-pumped quantum dots 
via their environmental phonon reservoir \cite{tema}, respectively. Related results can be achieved for two-spin qubits that interact 
via microwave photons in a superconducting cavity \cite{srin}. Finally, photon correlations are being establishing in architectures 
with two or more cavities interacting with a DQD \cite{hell,nian2}.

Motivated by the present progress in these directions, here we investigate the quantum dynamics of a leaking microwave resonator 
mode coupled in two-photon resonance with a semiconductor DQD qubit. Single electrons may tunnel from the source lead to the 
DQD states, which are coupled to the cavity mode, respectively, and then tunnel to the drain while the cavity mode accumulates 
photons. We have found a critical value for the qubit-cavity couplings such that the mean-number of the cavity photons substantially 
enhances or suppresses, by slightly varying the coupling strength around this value. The electric current through the DQD sample 
follows the cavity photon dynamics in the steady-state, suggesting a convenient way to convert electric current in a microwave 
photon flux. Furthermore, by measuring the current one can estimate the photon flux intensity, or vice versa. The phonons, 
presented in the system due to the involved materials features, may lower the resonator photon number, under the adopted 
approximations, whereas the electric current can be slighter bigger or lesser compared to the situations without considering 
phonon influences at low temperatures. Finally, the photon statistics changes within thermal-like to higher super-Poissonian 
features or Poissonian photon statistics, respectively, i.e., towards achieving single-qubit lasing effects at microwave frequency 
ranges.

This paper is organized as follows. In Sec.~\ref{theo} we describe the analytical approach and the system of interest, while 
in Sec.~\ref{seqm} we represent the equations of motion characterising the discussed system, respectively. Sec.~\ref{RD} 
presents and analyses the obtained results. The article concludes with a summary given in Sec.~\ref{sum}.

\section{Analytical approach \label{theo}}
We shall consider a hybrid setup comprising a semiconducting DQD qubit which is coupled to an electromagnetic microwave 
resonator as well as to source and drains leads, respectively. The Coulomb blockade regime is considered here meaning that 
the DQD is restricted to three possible configurations \cite{rev_art}: the null-electron subspace or the empty-dot state, i.e. 
$|o\rangle$, together with the single-electron subspaces, where the electron is localized either on the left, $|L\rangle$, or 
the right dot, $|R\rangle$, with $|o\rangle\langle o|+|L\rangle\langle L|+|R\rangle\langle R|=1$, respectively. Therefore, 
the electronic pumping process occurs as follows: an electron may tunnel from the source lead to the DQD states, i.e. 
$|L\rangle \to |R\rangle$, and then tunnels to the drain, correspondingly. Furthermore, we assume that the semiconductor 
DQD exchanges energy with the resonator mode when its generalized frequency $\Omega$ is equal approximately to the 
doubled value of the cavity frequency $\omega_{r}$. This is a complex non-linear system and solving it analitically in the 
steady-state is a challenging task. Therefore, we shall adapt the existing theoretical tools in order to be able to investigate 
it in certain cases of practical interest. Hence, the entire Hamiltonian describing the investigated system is given by 
\begin{eqnarray}
H=H_{q}+H_{r}+H_{qpn}+H_{rpt}. \label{Ht}
\end{eqnarray}
Here, see e.g. \cite{rev_art},
\begin{eqnarray}
H_{q}=\frac{\hbar\epsilon}{2}\sigma_{z} +\hbar\tau(\sigma^{+}+\sigma^{-}),
\label{Hq}
\end{eqnarray}
is the Hamiltonian of two quantum dots, forming the DQD qubit, with $\epsilon$ being their energy separation, while $\tau$ is the 
inter-dot tunnelling amplitude, both controlled experimentally, whereas
\begin{eqnarray}
H_{r}=\hbar\omega_{r}a^{\dagger}a + \hbar g\sigma_{z}(a^{\dagger}+a),
\label{Hr}
\end{eqnarray}
describes the microwave resonator's free energy and qubit-cavity interaction, respectively, with $g$ being the corresponding coupling strength.
The cavity photon creation (annihilation) operators, $a^{\dagger}(a)$, obey the standard commutation relations: $[a,a^{\dagger}]=1$,
$[a,a]=[a^{\dagger},a^{\dagger}]=0$. On the other side, the qubit's operators are defined as follows: $\sigma_{z}=|L\rangle\langle L| 
- |R\rangle\langle R|$, $\sigma^{+}=|L\rangle \langle R|$ and $\sigma^{-}=|R\rangle \langle L|$, which obey the commutation relations 
for su(2) algebra, that is, $[\sigma^{+},\sigma^{-}]=\sigma_{z}$ and $[\sigma_{z},\sigma^{\pm}]=\pm 2\sigma^{\pm}$. 
The electron-phonon interaction, mainly governed by the material's properties, is characterized by the next Hamiltonian, i.e.,
\begin{eqnarray}
H_{qpn}=\sum_{p}\hbar\omega_{p}b^{\dagger}_{p}b_{p} + \hbar\sigma_{z}\sum_{p}g_{p}(b^{\dagger}_{p}+b_{p}),
\label{Hfn}
\end{eqnarray}
where the first component is the free energy of the phonon reservoir, with $b^{\dagger}_{p}$ and $b_{p}$ being the generation and 
annihilation phonon operators satisfying the boson-commutation relations, $[b_{p},b^{\dagger}_{p'}]=\delta_{pp'}$ and 
$[b_{p},b_{p'}]=[b^{\dagger}_{p},b^{\dagger}_{p'}]=0$, while the second one accounts for the DQD-phonons interaction, respectively. 
The last Hamiltonian, i.e. $H_{rpt}$, has the following expression
\begin{eqnarray}
H_{rpt}=\sum_{k}\hbar\omega_{k}a^{\dagger}_{k}a_{k} + i\hbar\sum_{k}\chi_{k}(a^{\dagger} + a)(a^{\dagger}_{k}-a_{k}), \label{Hft}
\end{eqnarray}
and considers the interaction of the microwave resonator's single-mode with its corresponding thermal electromagnetic field (EMF) reservoir, 
described by the photon annihilation and creation operators $\{a_{k},a^{\dagger}_{k}\}$, satisfying the same bosonic commutation relations 
as those for phonon operators. Again, here, the first term corresponds to the free energy of the EMF reservoir, whereas the second one 
denotes the interaction of the resonator mode with its surrounding thermal bath. Notice that the DQD pumping, via the electronic reservoirs, 
will be introduced at the end of this Section.

Diagonalizing the Hamiltonian (\ref{Hq}), using the transformation 
\begin{eqnarray}
|L\rangle=\cos{(\theta/2)}|e\rangle-\sin{(\theta/2)}|g\rangle, \nonumber \\ 
|R\rangle=\sin{(\theta/2)}|e\rangle + \cos{(\theta/2)}|g\rangle, \label{dst} 
\end{eqnarray}
with $\cos{\theta}=\epsilon/\Omega$, $\sin{\theta}=2\tau/\Omega$ and $\Omega=\sqrt{\epsilon^{2}+(2\tau)^{2}}$, one arrives at a 
Hamiltonian $H$ represented via new qubit's quasispin operators: $R_{z}=|e\rangle\langle e| - |g\rangle\langle g|$, $R_{eg}=|e\rangle\langle g|$ 
and $R_{ge}=|g\rangle\langle e|$ which satisfy the commutation relations: $[R_{z},R_{eg}]=2R_{eg}$, $[R_{z},R_{ge}]=-2R_{ge}$, and 
$[R_{eg},R_{ge}]=R_{z}$, respectively. Then in the interaction picture, given by the unitary operator $U(t)=\exp(iH_{0}t/\hbar)$ with 
$H_{0}=\hbar \Omega R_{z}/2 + \hbar\omega_{r}a^{\dagger}a + \sum_{k}\hbar\omega_{k}a^{\dagger}_{k}a_{k} + 
\sum_{p}\hbar\omega_{p}b^{\dagger}_{p}b_{p}$, one obtains the following Hamiltonian: $H_{I}=H_{s}+\bar H_{f}$, where
\begin{eqnarray}
H_{s}&=& - \hbar\sin{\theta}\sum_{p}g_{p}R_{ge}b^{\dagger}_{p}e^{-i(\Omega-\omega_{p})t} \nonumber \\
&-& i\hbar\sum_{k}\chi_{k}a^{\dagger}a_{k}e^{-i(\omega_{k}-\omega_{r})t} + H.c.,
\label{Hs}
\end{eqnarray}
is the so-called slow part of the Hamiltonian, which will lead to the standard qubit's phonon decay and resonator's mode photon leaking effects, 
respectively, while
\begin{eqnarray}
\bar H_{f} = \sum_{\alpha \in \{0, \cdots, 5\}}\bar H^{(\alpha)}_{f}. \label{Hbf}
\end{eqnarray}
Here,
\begin{eqnarray}
\bar H^{(0)}_{f}&=& - \hbar\bar\delta_{0}R^{2}_{z}+\hbar\bar\delta R_{z}(1 +2a^{\dagger}a), \label{Hb0}
\end{eqnarray}
describes the frequency shifts of the DQD qubit's states which can be dependent on the resonator's photon number, while 
\begin{eqnarray*}
\bar \delta_{0}&=&g^{2}\bigl(\cos^{2}{\theta}-\omega_{r}^{2}\sin^{2}{\theta}/(\Omega_{-}\Omega_{+})\bigr)/\omega_{r}, \nonumber \\
\bar\delta&=&g^{2}\Omega\sin^{2}{\theta}/(\Omega_{-}\Omega_{+}), 
\end{eqnarray*}
with $\Omega_{\pm}=\Omega \pm \omega_{r}$ and $g/\omega_{r} \ll 1$. The next term of the Hamiltonian (\ref{Hbf}), i.e.,
\begin{eqnarray}
\bar H^{(1)}_{f}=\hbar\bar g\bigl(R_{eg}a^{2}e^{i(\Omega-2\omega_{r})t} + H.c. \bigr), \label{Hb1}
\end{eqnarray}
where 
\begin{eqnarray*}
\bar g=g^{2}\Omega\sin{2\theta}/(2\omega_{r}\Omega_{-}), 
\end{eqnarray*}
accounts for transitions among the qubit's energy states, i.e. 
$|e\rangle \leftrightarrow |g\rangle$, via two-photon processes. Generally, the Hamiltonians (\ref{Hb0}) and (\ref{Hb1}) characterise the 
cavity-qubit coherent evolution. However, the Hamiltonian
\begin{eqnarray}
\bar H^{(2)}_{f}&=&\sum_{p}g_{p}R_{ge}b^{\dagger}_{p}\bigl(g_{-}a^{\dagger}e^{i(\omega_{p} - \Omega_{-})t} 
+ g_{+}a e^{i(\omega_{p} - \Omega_{+})t}\bigr) \nonumber \\ 
&+& H.c., \label{Hb2}
\end{eqnarray}
with 
\begin{eqnarray*}
g_{\pm}=g\sin{2\theta}/\Omega_{\pm}, 
\end{eqnarray*}
describes reservoir phonon generation or annihilation processes at frequencies $\omega_{p}=\Omega_{\pm} \equiv \Omega \pm \omega_{r}$ 
accompanied by absorption or creation of a cavity photon, respectively, while the DQD qubit is making a $|e\rangle \to |g\rangle$ transition, or 
conversely. The spontaneous decay of the qubit is described by the Hamiltonian $\bar H^{(3)}_{f}$, i.e.,
\begin{eqnarray}
\bar H^{(3)}_{f}=- ig_{s}\sum_{k}\chi_{k}R_{ge}a^{\dagger}_{k}e^{i(\omega_{k}-\Omega)t} + H.c., \label{Hb3}
\end{eqnarray}
where 
\begin{eqnarray*}
g_{s}=g\sin{\theta}/\Omega_{+}, 
\end{eqnarray*}
while the phonon generation or annihilation processes followed by the absorption or creation of a cavity photon, while the qubit resides in the 
same state, are given respectively by the Hamiltonian $\bar H^{(4)}_{f}$, namely,
\begin{eqnarray}
\bar H^{(4)}_{f}=g_{z}R_{z}\sum_{p}g_{p}ab^{\dagger}_{p}e^{i(\omega_{p}-\omega_{r})t} + H.c.. \label{Hb4}
\end{eqnarray}
Here 
\begin{eqnarray*}
g_{z}=\sin{\theta}g_{s}. 
\end{eqnarray*}
Finally, two-phonon processes where their frequencies satisfy the relation $\omega_{p2}-\omega_{p1}=\Omega$ during qubit's transition 
$|e\rangle \leftrightarrow |g\rangle$ are characterized by the following Hamiltonian,
\begin{eqnarray}
\bar H^{(5)}_{f}&=&\sin{\theta}\sum_{p1\not=p2}\frac{g_{p1}g_{p2}}{\omega_{p2}}R_{eg}b^{\dagger}_{p1}b_{p2}
e^{i(\omega_{p1}+\Omega-\omega_{p2})t} \nonumber \\
&+& H.c., \label{Hb5}
\end{eqnarray}
see also \cite{ndqd1}, or complementarily \cite{ekm,toexp,multp}, for two- or multi-photon effects respectively. Notice that the Hamiltonian 
$(\ref{Hbf})$ was obtained via the relation: 
\begin{eqnarray}
\bar H_{f}=-\frac{i}{\hbar}H_{f}(t)\int dt H_{f}(t), \label{Hff}
\end{eqnarray}
see e.g. \cite{ttr1,ttr2}, where $H_{f}(t)$ is the corresponding fastly oscillating part of the entire Hamiltonian, containing the counter-rotating 
terms, after applying the unitary operator $U(t)$. Actually, the Hamiltonians (\ref{Hb2}-\ref{Hb5}) describe the cavity-DQD's two-quanta 
decay processes, respectively, when $g/\omega_{r} \ll 1$. In this way, we have arrived at an effective Hamiltonian, $H_{I}$=$H_{s}+\bar H_{f}$, 
in the interaction picture, characterizing both the involved linear and non-linear phenomena, occurring due to the two-photon nature of the 
qubit-resonator interaction and the interplay among the photon and phonon effects.

Next, assuming that the corresponding resonator's as well as DQD's interaction with the photon or phonon environmental reservoirs are weak, 
one can eliminate these degrees of freedom in the Born-Markov approximations \cite{gsag,kmek}. In this regard, one inserts separately each 
of the Hamiltonians from (\ref{Hs}) or (\ref{Hbf}) in the time-evolution equation for the density matrix operator $\rho$, i.e.,
\begin{eqnarray}
\dot\rho(t) =-\frac{1}{\hbar^{2}}\int^{t}_{0}dt'{\rm Tr_{r}}\bigl\{\bigl[H_{I}(t),[H_{I}(t'),\rho(t')]\bigr]\bigr\}, \label{eqms}
\end{eqnarray}
where the overdot means differentiation with respect to time, whereas the notation ${\rm Tr_{r}}\{ \cdots \}$ means the trace over the 
corresponding phonon or photon degrees of freedom. Afterwards, performing the trace and the Born-Markov approximations, one arrives at 
the following master equation describing {\it the qubit plus leaking resonator's mode} subsystem only,
\begin{widetext}
\begin{eqnarray}
\dot\rho(t) &+& \frac{i}{\hbar}[\bar H_{0},\rho]= 
- \frac{1}{2}\biggl\{\Gamma_{1}[R_{ge},R_{eg}\rho] + \Gamma_{2}[R_{eg},R_{ge}\rho]
+\Gamma_{+}\bigl\{(1+\bar n_{+})[R_{eg}a^{\dagger},aR_{ge}\rho] + \bar n_{+}[R_{ge}a,a^{\dagger}R_{eg}\rho]\bigr\}
\nonumber \\
&+& \Gamma_{-}\bigl\{(1+\bar n_{-})[R_{eg}a,a^{\dagger}R_{ge}\rho] + \bar n_{-}[R_{ge}a^{\dagger},aR_{eg}\rho]\bigr\}
+ \Gamma_{z}\bigl\{(1+\bar n)[R_{z}a^{\dagger},aR_{z}\rho] + \bar n[R_{z}a,a^{\dagger}R_{z}\rho]\bigr\} 
\nonumber \\
&+& \kappa\bigl\{(1+\bar n)[a^{\dagger},a\rho] + \bar n[a,a^{\dagger}\rho]\bigr\} + \gamma_{d}\cos^{2}{\theta}[R_{z},R_{z}\rho]
+\Gamma_{L}\bigl\{\cos^{2}{(\theta/2)}[R_{oe},R_{eo}\rho] + \sin^{2}{(\theta/2)}[R_{og},R_{go}\rho] \bigr \} \nonumber \\
&+&\Gamma_{R}\bigl\{\sin^{2}{(\theta/2)}[R_{eo},R_{oe}\rho] + \cos^{2}{(\theta/2)}[R_{go},R_{og}\rho]\bigr\} \biggr \} + H.c., 
\label{mqq}
\end{eqnarray}
\end{widetext}
where
\begin{eqnarray*}
\bar H_{0}=\hbar R_{z}(\Delta + 2\bar\delta a^{\dagger}a - \bar\delta_{0}R_{z}) + \hbar \bar g(R_{eg}a^{2} + R_{ge}a^{\dagger2}),
\end{eqnarray*}
and $\Delta=\bar\delta + (\Omega-2\omega_{r})/2$. The following notations were used there: $\Gamma_{1}=\bar n(\Omega)\bar \Gamma(\Omega)
+\gamma_{d}\sin^{2}{\theta}$, $\Gamma_{2}=\bigl(1+\bar n(\Omega)\bigr)\bar \Gamma(\Omega) + \gamma_{d}\sin^{2}{\theta}$, 
$\bar \Gamma(\Omega)=\sin^{2}{\theta}\Gamma(\Omega)$ + $g^{2}_{s}\kappa(\Omega)$, $\Gamma_{\pm}=
g^{2}_{\pm}\Gamma(\Omega_{\pm})$, $\Gamma_{z} = g^{2}_{z}\Gamma(\omega_{r})$, $\bar n_{\pm}\equiv \bar n(\Omega_{\pm})$, 
$\bar n \equiv \bar n(\omega_{r})$, and $\kappa \equiv \kappa(\omega_{r})$, with $\bar n(\omega)=1/\bigl[\exp{(\hbar \omega/k_{B}T)}-1\bigr]$ 
being the mean photon or phonon numbers at the frequency $\omega$ and temperature $T$, while $k_{B}$ is the Boltzmann constant. 
Further, $\Gamma(\omega)=\pi\sum_{p}g^{2}_{p}\delta(\omega_{p} - \omega)$ denotes the phonon decay rate, whereas 
$\kappa(\omega)=\pi\sum_{k}\chi^{2}_{k}\delta(\omega_{k}-\omega)$ represents the cavity photon decay rate, both at the frequency 
$\omega$, respectively. Note, here, that in the final master equation (\ref{mqq}) we have omitted the two-phonon effects, described by 
the Hamiltonian (\ref{Hb5}), because we are interested in lower temperatures limits, where this process is weak. Additionally, we have 
added in the usual way at weaker temperatures, the dephasing terms proportional to $\gamma_{d}$ as well as the corresponding 
electronic pumping of the state $|L\rangle$, proportional to the tunnelling rate $\Gamma_{L}$, together with processes of unloading 
the state $|R\rangle$, given by the tunnelling rate $\Gamma_{R}$, see e.g. \cite{rev_art,lamb}, that is,
\begin{eqnarray}
\Lambda_{d}\rho &=& -\frac{\gamma_{d}}{2}\bigl\{[\sigma_{z},\sigma_{z}\rho] + H.c. \bigr\}, \nonumber \\
\Lambda_{L}\rho &=& -\frac{\Gamma_{L}}{2}\bigl \{\bigl[|o\rangle \langle L|,|L\rangle \langle 0|\rho\bigr] + H.c.\bigr\}, \nonumber \\
\Lambda_{R}\rho &=& -\frac{\Gamma_{R}}{2}\bigl \{\bigl[|R\rangle \langle o|,|o\rangle \langle R|\rho\bigr] + H.c.\bigr\}. 
\label{rdmp}
\end{eqnarray}
In Eq.~(\ref{mqq}), these contributions are given in the secular approximation, after use of transformation (\ref{dst}), meaning that 
$\Omega \gg \{\Gamma_{L/R},\gamma_{d}\}$ in our approach. The secular approximation was used in deriving the master equation
Eq.~(\ref{mqq}) too, which is valid if the qubit's frequency $\Omega$ is larger than any decay rate in the system or coupling strengths 
\cite{kmek}, respectively.

In the following section, we shall make use of Eq.~(\ref{mqq}) in order to obtain the corresponding equations of motion describing 
the quantum dynamics of the combined resonator-DQD subsystem.
\section{The equations of motion \label{seqm}}
Using the master equation (\ref{mqq}), one can obtain the following exact system of equations of motion describing the entire sample 
incorporating a semiconductor DQD qubit coupled respectively in two-photon resonance to the cavity boson-mode, i.e.,
\begin{eqnarray}
\dot P^{(0)}_{n}&=& -\Gamma^{(0)}_{n}P^{(0)}_{n} + \Gamma_{R}\cos^{2}(\theta/2)P^{(1)}_{n} + \Gamma_{R}\sin^{2}(\theta/2)P^{(2)}_{n} 
 \nonumber \\
&+& \kappa\bar n n P^{(0)}_{n-1}+ \kappa(1+\bar n)(1+n)P^{(0)}_{n+1}, \nonumber \\
\dot P^{(1)}_{n} &=& - \Gamma^{(1)}_{n}P^{(1)}_{n} + \Gamma_{2}P^{(2)}_{n} + \Gamma_{+}(1+\bar n_{+})(1+n)P^{(2)}_{n+1}  
\nonumber \\
&+& \Gamma_{-}(1+\bar n_{-})nP^{(2)}_{n-1} -i\bar g P^{(5)}_{n} + \Gamma_{L}\sin^{2}(\theta/2)P^{(0)}_{n} \nonumber \\
&+& (\Gamma_{z}+\kappa)\bigl((1+\bar n)(1+n)P^{(1)}_{n+1} + \bar n n P^{(1)}_{n-1}\bigr), \nonumber \\
\dot P^{(2)}_{n} &=& - \Gamma^{(2)}_{n}P^{(2)}_{n} + (\Gamma_{z}+\kappa)\bigl((1+\bar n)(1+n)P^{(2)}_{n+1} \nonumber \\
&+& \bar n n P^{(2)}_{n-1}\bigr) + \Gamma_{1}P^{(1)}_{n} +\Gamma_{+}\bar n_{+}nP^{(1)}_{n-1} - i\bar g P^{(3)}_{n}\nonumber \\
&+&\Gamma_{-}\bar n_{-}(1+n)P^{(1)}_{n+1} + \Gamma_{L}\cos^{2}(\theta/2)P^{(0)}_{n}, \nonumber \\
\dot P^{(3)}_{n} &=& - \Gamma^{(3)}_{n}P^{(3)}_{n} - (\Gamma_{z}-\kappa)\bigl((1+\bar n)(1+n)P^{(3)}_{n+1} \nonumber \\
&-& \bar n(n+2)P^{(7)}_{n}\bigr) + 2i\bigl(\Delta + 2\delta(n+1)\bigr)P^{(4)}_{n}  \nonumber \\
&-& 2i\bar g(n+1)(n+2)\bigl(P^{(2)}_{n}-P^{(1)}_{n+2}\bigr), \nonumber \\
\dot P^{(4)}_{n} &=& - \Gamma^{(4)}_{n}P^{(4)}_{n} - (\Gamma_{z}-\kappa)\bigl((1+\bar n)(1+n)P^{(4)}_{n+1} \nonumber \\
&+& \bar n(n+2)P^{(8)}_{n}\bigr) + 2i\bigl(\Delta + 2\delta(n+1)\bigr)P^{(3)}_{n},  \nonumber \\
\dot P^{(5)}_{n} &=& - \Gamma^{(5)}_{n}P^{(5)}_{n} - (\Gamma_{z}-\kappa)\bigl(\bar n n P^{(5)}_{n-1} +
 (1+\bar n)(n-1) \nonumber \\
&\times&P^{(7)}_{n}\bigr) - 2i\bigl(\Delta + 2\delta(n-1)\bigr)P^{(6)}_{n} - 2i\bar g n(n-1)\nonumber \\
&\times&\bigl(P^{(1)}_{n} - P^{(2)}_{n-2}\bigr), \nonumber \\
\dot P^{(6)}_{n} &=& - \Gamma^{(6)}_{n}P^{(6)}_{n} - (\Gamma_{z}-\kappa)\bigl((1+\bar n)(n-1)P^{(8)}_{n} \nonumber \\
&+& \bar n nP^{(6)}_{n-1}\bigr) - 2i\bigl(\Delta + 2\delta(n-1)\bigr)P^{(5)}_{n},  \nonumber \\
\dot P^{(7)}_{n} &=& - \Gamma^{(7)}_{n}P^{(7)}_{n} - 2i\bigl(\Delta + 2\delta n\bigr)P^{(8)}_{n} - 2i\bar g n(n+1) \nonumber \\
&\times&\bigl(P^{(1)}_{n+1} - P^{(2)}_{n-1}\bigr) + (\Gamma_{z}-\kappa)\bigl((\bar n+1)nP^{(3)}_{n} \nonumber \\
&-& \bar n(n+1)P^{(5)}_{n}\bigr), \nonumber \\
\dot P^{(8)}_{n} &=& - \Gamma^{(8)}_{n}P^{(8)}_{n} - 2i\bigl(\Delta+2\delta n)\bigr)P^{(7)}_{n} - (\Gamma_{z}-\kappa) \nonumber \\
&\times&\bigl(n(1+\bar n)P^{(4)}_{n} + \bar n(1+n)P^{(6)}_{n}\bigr). 
\label{eqm}
\end{eqnarray}
The system of equations (\ref{eqm}) can be easily obtained using the master equation (\ref{mqq}), if one first gets the equations of motion 
for $\rho_{\alpha\beta}$=$\langle \alpha|\rho|\beta \rangle$, $\alpha,\beta \in \{o, e,g\}$, see also \cite{ttm}, and then writing the 
corresponding equations for the following variables: $\rho^{(0)}=\rho_{oo}$, $\rho^{(1)}=\rho_{gg}$, $\rho^{(2)}=\rho_{ee}$, 
$\rho^{(3)}=a^{2}\rho_{ge}-\rho_{eg}a^{\dagger 2}$, $\rho^{(4)}=a^{2}\rho_{ge}+\rho_{eg}a^{\dagger 2}$, $\rho^{(5)}= 
a^{\dagger 2}\rho_{eg}- \rho_{ge}a^{2}$, $\rho^{(6)}=a^{\dagger 2}\rho_{eg} + \rho_{ge}a^{2}$, $\rho^{(7)} = 
a^{\dagger}\rho_{eg}a^{\dagger}-a\rho_{ge}a$ and $\rho^{(8)} = a^{\dagger}\rho_{eg}a^{\dagger} + a\rho_{ge}a$. The 
projection on the Fock states $|n\rangle$, i.e. $P^{(j)}_{n}=\langle n|\rho^{(j)}|n\rangle$, with $j \in \{0, \cdots, 8\}$ and 
$n \in \{0,\infty\}$, will lead us to the system of equations (\ref{eqm}). The corresponding decay rates, i.e. $\Gamma^{(j)}_{n}$, 
$j \in \{0, \cdots, 8\}$, are given in Appendix \ref{apxA}. Notice also that in many studies on this topic the corresponding equations 
of motion are obtained by decoupling the qubit-cavity correlators, i.e., $\langle \hat Q\hat F\rangle \approx \langle \hat Q\rangle
\langle \hat F\rangle$, see e.g. \cite{kul,gul,sql2,sql3}, an approximation valid if the corresponding operator's quantum fluctuations 
are negligible. In our approach, however, this approximation is avoided.
\begin{figure}[t]
\includegraphics[width =7cm]{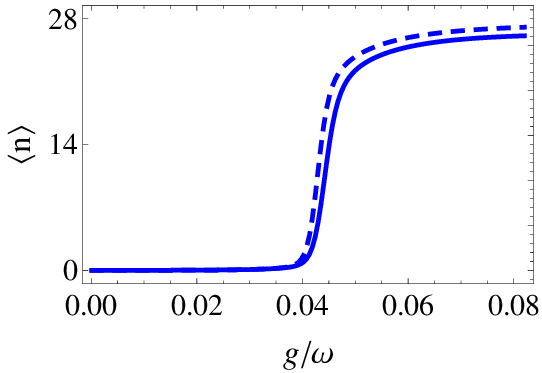}
\caption{\label{fig-1}
The steady-state behaviour of the cavity mean photon number $\langle n\rangle=\langle a^{\dagger}a\rangle$ as a function of the ratio of 
the qubit-resonator coupling strength $g$ over the cavity frequency $\omega_{r}$. Here $\tau/\epsilon=0.1$, $\epsilon/\omega_{r}=1.96$, 
$\hbar\omega_{r}/(k_{B}T)=5$, $\Gamma_{L}/\omega_{r}=\Gamma_{R}/\omega_{r}=0.03$, $\gamma_{d}/\omega_{r}=10^{-3}$, 
$\kappa/\omega_{r}=7\cdot 10^{-4}$ and $\kappa(\Omega)/\omega_{r}=10^{-4}$. The solid line corresponds to
$\Gamma(\Omega)/\omega_{r}=0.02$, $\Gamma(\Omega_{\pm})/\omega_{r}=0.01$, and 
$\Gamma(\omega_{r})/\omega_{r}=0.01$, whereas the dashed one to $\Gamma(\Omega)$=$\Gamma(\Omega_{\pm})$=
$\Gamma(\omega_{r})=0$, i.e., no phonons involved.}
\end{figure}

Generally, to solve the infinite system of equations (\ref{eqm}), one truncates it at a certain maximum value $n=n_{max}$ so that a further 
increase in its value, i.e. $n_{max}$, does not modify the obtained results. As a consequence, the steady-state cavity-mode photon mean 
number, i.e. $\langle n\rangle =\langle a^{\dagger}a\rangle$, is expressed as:
\begin{eqnarray}
\langle n\rangle = \sum^{n_{max}}_{n=0}n\bigl(P^{(0)}_{n}+ P^{(1)}_{n}+P^{(2)}_{n}\bigr), \label{nmf}
\end{eqnarray}
with
\begin{eqnarray*}
\sum^{n_{max}}_{n=0}\bigl(P^{(0)}_{n}+ P^{(1)}_{n}+P^{(2)}_{n}\bigr)=1.
\end{eqnarray*}
The corresponding steady-state second-order cavity photon correlation function is defined in the usual way \cite{glb}, namely,
\begin{eqnarray}
g^{(2)}(0)&=&\frac{\langle a^{\dagger 2}a^{2}\rangle}{\langle n\rangle^{2}} \label{crf} \\
&=&\frac{1}{\langle n\rangle^{2}}\sum^{n_{max}}_{n=0}n(n-1)\bigl(P^{(0)}_{n}+ P^{(1)}_{n}+P^{(2)}_{n}\bigr). \nonumber 
\end{eqnarray}
Note that $g^{(2)}(0)<1$ means sub-Poissonian, $g^{(2)}(0)>1$ super-Poissonian, and $g^{(2)}(0)=1$ Poissonian photon statistics, 
respectively.

The current $I_{q}$ through the DQD qubit is proportional to the population in the state $|R\rangle$ \cite{xu}, that is, 
\begin{eqnarray}
I_{q} &=& |e|\Gamma_{R}|R\rangle \langle R|  \label{Iqq} \\
&=& |e|\Gamma_{R}\sum^{n_{max}}_{n=0}\bigl(\sin^{2}{(\theta/2)}P^{(2)}_{n} + \cos^{2}{(\theta/2)}P^{(1)}_{n}\bigr), 
\nonumber 
\end{eqnarray}
where $e$ is the electron's charge. In Exp.~(\ref{Iqq}), we have applied the transformation (\ref{dst}) and the secular approximation. 
\begin{figure}[t]
\includegraphics[width =7cm]{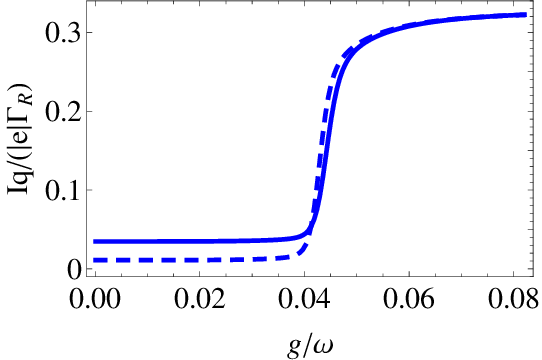}
\caption{\label{fig-2}
The steady-state dependence of the current $I_{q}$ through the DQD system as a function of $g/\omega_{r}$. 
All other involved parameters are the same as in Fig.~(\ref{fig-1}).}
\end{figure}

In the following Section, we shall describe the photon resonator's quantum dynamics as well as the steady-state behaviours of the electrical current, 
based on results obtained from Eqs.~(\ref{eqm}-\ref{Iqq}).

\section{Results and discussion \label{RD}}
Once we have obtained the corresponding equations of motion, i.e. Eqs.~(\ref{eqm}), in Fig.~(\ref{fig-1}) we show the steady-state 
dependences of the mean cavity photon number as a function of the DQD-resonator coupling strength, near resonance, i.e. 
$\Omega \approx 2\omega_{r}$. We have considered a weakly leaking resonator having the frequency $\omega_{r} \sim 1{\rm GHz}$ 
embedded in a ${\rm mK}$ temperature environment, so that $\hbar\omega_{r}/(k_{B}T)=5$. Generally, the decay rates were selected 
within the secular approximation, that is, smaller than qubit's frequency $\Omega$, in concordance with the valability of the master equation 
(\ref{mqq}). Particularly, we observe a threshold transition when the mean-photon number abruptly changes from lower to higher values, 
or conversely, by slightly varying the qubit-resonator coupling strength, see also Ref.~\cite{sql2} for related single-photon processes. This 
behaviour is typical for an externally coherently pumped and leaking cavity mode, containing a Kerr-like non-linearity \cite{drm,mmk}, 
for instance. Here, however, this comporting is due to the unilateral electronic pumping of the DQD sample, i.e. 
$|o\rangle \to |L\rangle \to |R\rangle \to|o\rangle$. The presence of phonons in the sample, lower the mean resonator's photon 
number above threshold, compare the solid and dashed curves in Fig.~(\ref{fig-1}).
\begin{figure}[t]
\includegraphics[width =7cm]{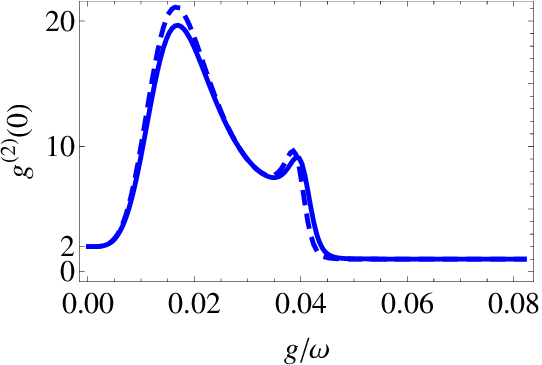}
\caption{\label{fig-3}
The normalized steady-state second-order photon-photon correlation function $g^{(2)}(0)$ versus the ratio $g/\omega_{r}$. 
Other parameters are as in Fig.~(\ref{fig-1}).}
\end{figure}

On the other side, Fig.~(\ref{fig-2}) depicts the steady-state current through the DQD qubit which is proportional to the population 
in the state $|R\rangle$. It follows the same behaviour as the mean photon number, see Fig.~(\ref{fig-2}) and Fig.~(\ref{fig-1}). 
Therefore, measuring the current through the DQD one can estimate the photon flux leaking out from the cavity mode, or vice 
versa. The current may enhance or diminuish in the presence or the absence of phonons with respect to the critical coupling 
strength, respectively. The discrepancies rely on the fact that phonons may contribute to population on the right state $|R\rangle$, 
but not necessarily to the exchange of photons with the cavity mode. Practically, the described process is a useful tool to convert 
electric current into microwave photons in DQD-cavity systems. Moreover, the cavity photons obey the super-Poissonian statistics 
below the threshold, i.e. $g^{(2)}(0) \ge 2$, see Fig.~(\ref{fig-3}), which changes to Poissonian photons statistics above the 
threshold, i.e. to microwave photon lasing effects where $g^{(2)}(0)=1$, respectively, see also \cite{mas1,gul,hartk,sql2}. 
Furthermore, around threshold, the photons correlation evidently enhances, while in the absence of environmental thermal 
photons, i.e. when $T \to 0$, $g^{(2)}(0)$ does not converge for $g/\omega_{r}=0$, so one should proceed with non-zero 
values for DQD-cavity coupling strengths in this particular case.

\section{Summary \label{sum}}
Summarizing, we have investigated a semiconducting DQD qubit coupled in two-photon resonance with a weakly leaking microwave resonator. 
The pumping of the qubit is achieved via the environmental electronic reservoirs in a way that always single electrons tunnel from the source 
lead to the DQD states, which interacts also with the cavity mode, followed then by tunnelling to the drain, correspondingly. We have calculated 
the mean photon cavity number, under the secular approximation, and have found a critical value for the qubit-resonator coupling strength 
where the steady-state photon number changes abruptly, while the corresponding photon statistics vary from thermal-like, super-Poissonian 
to Poissonian photon statistics. The electric current through the DQD system follows the cavity-mode steady-state photon's number behaviours. 
This way, one achieves lasing effects in microwave frequency domains, while converting electric current in a photon flux. Respectively, one can 
estimate the intensity of the photon flux, generated in the cavity mode, via measuring the electric current, or conversely. Finally, the phonons 
presented in the sample, slightly modify the presented results within the secular approximation, i.e. $\Omega \gg \Gamma_{R/L} > 
\{\Gamma(\Omega), \Gamma(\omega_{\pm}),\Gamma(\omega_{r}),\gamma_{d}\}$, and lower temperature as well as good-cavity limits,
respectively.

\acknowledgments
We highly appreciate the financial support from the Romanian Ministry of Research, Innovation and Digitization via the 5ROMD/20/05/2024/ 
research grant. MM is grateful for partial support from the Moldavian Ministry of Education and Research, through grant No. 011205, as well as 
for the nice hospitality of the Department of Theoretical Physics at the Horia Hulubei National Institute of Physics and Nuclear Engineering. 

\appendix
\section{The decay rates entering the equations of motion (\ref{eqm}) \label{apxA}}
Below one can find the corresponding decay rates which enter equation (\ref{eqm}), that is,
\begin{eqnarray}
\Gamma^{(0)}_{n}&=&\Gamma_{L} + \kappa n(1+\bar n) + \kappa\bar n(1+n), \nonumber \\
\Gamma^{(1)}_{n}&=&\Gamma_{1} + \Gamma_{R}\cos^{2}(\theta/2) + \Gamma_{+}\bar n_{+}(1+n) + \Gamma_{-}\bar n_{-}n 
\nonumber \\
&+& (\kappa +\Gamma_{z})\bigl(\bar n(1+n)+n(1+\bar n)\bigr), \nonumber \\
\Gamma^{(2)}_{n}&=&\Gamma_{2} + \Gamma_{R}\sin^{2}(\theta/2) + \Gamma_{+}n(1+\bar n_{+}) + \Gamma_{-}(1+\bar n_{-})
\nonumber \\
&\times&(1+n) + (\kappa +\Gamma_{z})\bigl(\bar n(1+n)+n(1+\bar n)\bigr), \nonumber \\
\Gamma^{(3)}_{n}&=&\Gamma_{\perp} + \Gamma_{+}\bigl(\bar n_{+}(n+3)+n(1+\bar n_{+})\bigr)/2 \nonumber \\
&+& \Gamma_{-}\bigl(\bar n_{-}(n+2)+ (1+n)(1+\bar n_{-})\bigr)/2 \nonumber \\
&+& (\kappa +\Gamma_{z})\bigl((1+\bar n)(1+n)+\bar n(n+2)\bigr), \nonumber \\
\Gamma^{(4)}_{n}&=&\Gamma^{(3)}_{n}, \nonumber \\
\Gamma^{(5)}_{n}&=&\Gamma_{\perp} + \Gamma_{+}\bigl(\bar n_{+}(n+1)+(n-2)(1+\bar n_{+})\bigr)/2 \nonumber \\
&+& \Gamma_{-}\bigl(\bar n_{-}n+ (n-1)(1+\bar n_{-})\bigr)/2 + (\kappa +\Gamma_{z}) \nonumber \\
&\times&\bigl((1+\bar n)(n-1)+n\bar n\bigr), \nonumber \\
\Gamma^{(6)}_{n}&=&\Gamma^{(5)}_{n}, \nonumber \\
\Gamma^{(7)}_{n}&=&\Gamma_{\perp} + \Gamma_{+}\bigl(\bar n_{+}(n+2)+(n-1)(1+\bar n_{+})\bigr)/2 \nonumber \\
&+& \Gamma_{-}\bigl(\bar n_{-}(1+n)+ n(1+\bar n_{-})\bigr)/2 + (\kappa +\Gamma_{z}) \nonumber \\
&\times&\bigl((1+\bar n)n + (1+n)\bar n\bigr), \nonumber \\
\Gamma^{(8)}_{n}&=&\Gamma^{(7)}_{n}, \label{dcr}
\end{eqnarray}
where $\Gamma_{\perp}=\bigl(\Gamma_{R} + \Gamma_{1} + \Gamma_{2}\bigr)/2$ + $2\gamma_{d}\cos^{2}{\theta}$.



\begin{thebibliography}{71}
\bibitem{expp1} A. Wallraff, D. I. Schuster, A. Blais, L. Frunzio, R.- S. Huang, J. Majer, S. Kumar, S. M. Girvin, and R. J. Schoelkopf,
Strong coupling of a single photon to a superconducting qubit using circuit quantum electrodynamics, Nature {\bf 431}, 162 (2004).

\bibitem{expp2} J. P. Reithmaier, G. Sek, A. L\"{o}ffler, C. Hofmann, S. Kuhn, S. Reitzenstein, L. V. Keldysh, V. D. Kulakovskii, T. L. Reinecke,
and A. Forchel, Strong coupling in a single quantum dot–semiconductor microcavity system, Nature {\bf 432}, 197 (2004).

\bibitem{theor} L. Childress, A. S. Sorensen, and M. D. Lukin, Mesoscopic cavity quantum electrodynamics with quantum dots, 
Phys. Rev. A {\bf 69}, 042302 (2004).

\bibitem{stokk} A. Stockklauser, P. Scarlino, J. V. Koski, S. Gasparinetti, C. K. Andersen, C. Reichl, W. Wegscheider,
T. Ihn, K. Ensslin, and A. Wallraff, Strong Coupling Cavity QED with Gate-Defined Double Quantum Dots Enabled by 
a High Impedance Resonator, Phys. Rev. X {\bf 7}, 011030 (2017).

\bibitem{scar} P.  Scarlino, D. J. van Woerkom, A. Stockklauser, J. V. Koski, M. C. Collodo, S. Gasparinetti, C. Reichl, W. Wegscheider, 
T. Ihn, K. Ensslin, and A. Wallraff, All-Microwave Control and Dispersive Readout of Gate-Defined Quantum Dot Qubits in Circuit Quantum 
Electrodynamics, Phys. Rev. Lett. {\bf 122}, 206802 (2019).

\bibitem{khan}  W.  Khan,  P. P. Potts, S. Lehmann, C. Thelander, K. A. Dick, P. Samuelsson, and V. F. Maisi, Efficient and continuous 
microwave photoconversion in hybrid cavity-semiconductor nanowire double quantum dot diodes, Nat. Commun. {\bf 12}, 5130 (2021).

\bibitem{stcp1} S.-S. Gu, S. Kohler, Y.-Q. Xu, R. Wu, Sh.-L. Jiang, Sh.-K. Ye, T. Lin, B.-Ch. Wang, and H.-O. Li, 
Probing Two Driven Double Quantum Dots Strongly Coupled to a Cavity, Phys. Rev. Lett. {\bf 130}, 233602 (2023).

\bibitem{stcp2} J. H. Ungerer, A. Pally, A. Kononov, S. Lehmann, J. Ridderbos, P. P. Potts, C. Thelander, K. A. Dick, V. F. Maisi, P. Scarlino,
A. Baumgartner, and C. Sch\"{o}nenberger, Strong coupling between a microwave photon and a singlet-triplet qubit,
Nat. Commun. {\bf 15}, 1068 (2024).

\bibitem{liuI} Y.-Y. Liu, K. D. Petersson, J. Stehlik, J. M. Taylor, and J. R. Petta, 
Photon Emission from a Cavity-Coupled Double Quantum Dot, Phys. Rev. Lett. {\bf 113}, 036801 (2014).

\bibitem{stock} A. Stockklauser, V. F. Maisi, J. Basset, K. Cujia, C. Reichl, W. Wegscheider, T. Ihn, A. Wallraff, and K. Ensslin,
Microwave Emission from Hybridized States in a Semiconductor Charge Qubit, Phys. Rev. Lett.  {\bf 115}, 046802 (2015).

\bibitem{haldar} S. Haldar,  H. Havir,  W. Khan, S. Lehmann, C. Thelander, K. A. Dick, and V. F. Maisi, 
Energetics of Microwaves Probed by Double Quantum Dot Absorption, Phys. Rev. Lett. {\bf 130}, 087003 (2023).

\bibitem{hhaldar} S. Haldar, D. Zenelaj, P. P. Potts, H. Havir, S. Lehmann, K. A. Dick, P. Samuelsson, and V. F. Maisi, 
Microwave power harvesting using resonator-coupled double quantum dot photodiode, 
Phys. Rev. B {\bf 109}, L081403 (2024).

\bibitem{hhhaldar} S.  Haldar, D. Barker, H. Havir, A. Ranni, S. Lehmann, K. A. Dick, and V. F. Maisi, Continuous Microwave 
Photon Counting by Semiconductor-Superconductor Hybrids, Phys. Rev. Lett. {\bf 133}, 217001 (2024).

\bibitem{stanis} O. Stanisavljevi\'{c}, J.-C. Philippe, J. Gabelli, M. Aprili, J. Est\'{e}ve, and J. Basset, Efficient Microwave 
Photon-to-Electron Conversion in a High-Impedance Quantum Circuit, Phys. Rev. Lett. {\bf 133}, 076302 (2024).

\bibitem{bur} L. E. Bruhat, T. Cubaynes, J. J. Viennot, M. C. Dartiailh, M. M. Desjardins, A. Cottet, and T. Kontos, 
Circuit QED with a quantum-dot charge qubit dressed by Cooper pairs, Phys. Rev. B {\bf 98}, 155313 (2018).

\bibitem{burk} G. Burkard, M. J. Gullans, X. Mi, and J. R. Petta, Superconductor–semiconductor hybrid-circuit quantum electrodynamics, 
Nat. Rev. Phys. {\bf 2}, 129–140 (2020).

\bibitem{havir} H. Havir, S. Haldar, W.  Khan, S. Lehmann, K. A. Dick, C. Thelander, P. Samuelsson, and V. F. Maisi, 
Quantum dot source-drain transport response at microwave frequencies, Phys. Rev. B {\bf 108}, 205417 (2023).

\bibitem{kul} M. Kulkarni, O. Cotlet, and H. E. T\"{u}reci, Cavity-coupled double-quantum dot at finite bias: 
analogy with lasers and beyond, Phys. Rev. B {\bf 90}, 125402 (2014).

\bibitem{rev_mod} A.J. Leggett, S. Chakravarty, A.T. Dorsey, M.P.A. Fisher, A. Garg, and W. Zwerger, 
Dynamics of the dissipative two-state system, Rev. Mod. Phys. {\bf 59}, 1 (1987).

\bibitem{rev_art} T. Brandes, Coherent and collective quantum optical effects in mesoscopic systems, 
Phys. Rep. {\bf 408}, 315 (2005).

\bibitem{elph1} R. S\'{a}nchez, G. Platero, and T. Brandes, Resonance Fluorescence in Transport through Quantum Dots: 
Noise Properties, Phys. Rev. Lett. {\bf 98}, 146805 (2007).

\bibitem{elph2} R. S\'{a}nchez, G. Platero, and T. Brandes, Resonance fluorescence in driven quantum dots: 
Electron and photon correlations, Phys. Rev. B  {\bf 78}, 125308 (2008).

\bibitem{agr2} B. K. Agarwalla, M. Kulkarni, S. Mukamel, and D. Segal, Tunable photonic cavity coupled to a 
voltage-biased double quantum dot system: Diagrammatic nonequilibrium Green’s function approach, 
Phys. Rev. B {\bf 94}, 035434 (2016).

\bibitem{chen} C.-C. Chen, T. M. Stace, and H.-S. Goan, Full-polaron master equation approach to dynamical steady states 
of a driven two-level system beyond the weak system-environment coupling, Phys. Rev. B {\bf 102}, 035306 (2020).

\bibitem{ghirri} A. Ghirri,  S. Cornia, and M. Affronte, Microwave photon detectors based on semiconducting double quantum dots, 
Sensors {\bf 20}, 4010 (2020).

\bibitem{nian} L.-L. Nian,  S. Hu, L. Xiong, J.-T. L\"{u}, and B. Zheng, Photon-assisted electron transport across a quantum phase 
transition, Phys. Rev. B {\bf 108}, 085430 (2023).

\bibitem{tarak} S. K. Hazra, L. Addepalli, P. K. Pathak, and T. N. Dey, 
Nondegenerate two-photon lasing in a single quantum dot, Phys. Rev. B {\bf 109}, 155428 (2024).

\bibitem{wong} C. H. Wong, and M. G. Vavilov, Quantum efficiency of a single microwave photon detector based on a semiconductor 
double quantum dot, Phys. Rev. A {\bf 95}, 012325 (2017).

\bibitem{xu} C.  Xu, and M. G. Vavilov,  Full counting statistics of photons emitted by a double quantum dot, 
Phys. Rev. B {\bf 88}, 195307 (2013).

\bibitem{zenelaj} D.  Zenelaj, P. P. Potts, and P. Samuelsson,  Full counting statistics of the photocurrent through a double quantum 
dot embedded in a driven microwave resonator, Phys. Rev. B {\bf 106}, 205135 (2022). 

\bibitem{wang} J.-Y.  Wang, L.-L. Nian, and J.-T.  L\"{u},  Engineering Quantum Criticality for Quantum Dot Power Harvesting, 
Chinese Phys. Lett. {\bf 41}, 020503 (2024).
    
\bibitem{zenelaj2} D. Zenelaj,  P. Samuelsson, and P. P.  Potts,  Wigner-function formalism for the detection of single microwave 
pulses in a resonator-coupled double quantum dot, 	arXiv:2410.14278 (2024). 

\bibitem{gu}  X. Gu, A.F. Kockum, A. Miranowicz, Y.X. Liu, and F. Nori, Microwave photonics with superconducting quantum circuits. 
Phys. Rep. {\bf 718–719}, 1 (2017).

\bibitem{gul3} M. J.  Gullans, J. Stehlik, Y.-Y. Liu, C. Eichler, J. R. Petta, and J. M. Taylor,  
Sisyphus Thermalization of Photons in a Cavity-Coupled Double Quantum Dot, 
Phys. Rev. Lett. {\bf 117}, 056801 (2016).

\bibitem{nian2} L.-L. Nian, B. Zheng, and J.-T. L\"{u}, Electrically driven photon statistics engineering in quantum-dot circuit 
quantum electrodynamics, Phys. Rev. B {\bf 107}, L241405 (2023).

\bibitem{nian3} L.-L. Nian, T.  Wang, and J.-T. L\"{u},  Plasmon Squeezing in Single-Molecule Junctions, 
Nano Lett. {\bf 22}, 9418-9423 (2022).

\bibitem{mas1} Y.-Y. Liu, J. Stehlik, C. Eichler, M. J. Gullans, J. M. Taylor, and J. R. Petta, Semiconductor double quantum dot micromaser,
Science {\bf 347}, 285 (2015).

\bibitem{gul} M. J. Gullans, Y.-Y. Liu, J. Stehlik, J. R. Petta, and J. M. Taylor, Phonon-Assisted Gain in a Semiconductor Double 
Quantum Dot Maser, Phys. Rev. Lett. {\bf 114}, 196802 (2015).

\bibitem{hartk} T. R. Hartke, Y.-Y. Liu, M. J. Gullans, and J. R. Petta, Microwave Detection of Electron-Phonon Interactions 
in a Cavity-Coupled Double Quantum Dot, Phys. Rev. Lett. {\bf 120}, 097701 (2018).

\bibitem{sql2} Y.-Y. Liu, J. Stehlik, C. Eichler, X. Mi, T. R. Hartke, M. J. Gullans, J. M. Taylor, and J. R. Petta,
Threshold Dynamics of a Semiconductor Single Atom Maser, Phys. Rev. Lett. {\bf 119}, 097702 (2017).

\bibitem{sql3} G. Rastelli, and M. Governale, Single atom laser in normal-superconductor quantum dots,
Phys. Rev. B {\bf 100}, 085435 (2019).

\bibitem{chlo} T.  Chlouba, and T. Novotn\'{y},  On the lack of intrinsic bistability of photon emission in a double quantum dot micromaser, 
J. Stat. Mech. {\bf 10}, 104009 (2019).

\bibitem{jin} P.-Q.  Jin, M. Marthaler, J. H. Cole, A. Shnirman, and G. Sch\"{o}n, Lasing and transport in a quantum-dot resonator circuit, 
Phys. Rev. B {\bf 84}, 035322 (2011).

\bibitem{lamb}  N. Lambert,  F.  Nori, and C. Flindt, Bistable Photon Emission from a Solid-State Single-Atom Laser, 
Phys. Rev. Lett. {\bf 115}, 216803 (2015).

\bibitem{rastM} M. Mantovani, A. D. Armour, W. Belzig, and G. Rastelli, Dynamical multistability in a quantum-dot laser,
Phys. Rev. B {\bf 99}, 045442 (2019).

\bibitem{agr1} B. K. Agarwalla, M. Kulkarni, and D. Segal, Photon statistics of a double quantum dot micromaser: 
Quantum treatment, Phys. Rev. B {\bf 100}, 035412 (2019).

\bibitem{taba} S. M.  Tabatabaei, and N. Jahangiri,  Lasing in a coupled hybrid double quantum dot-resonator system, 
Phys. Rev. B {\bf 101}, 115135 (2020).

\bibitem{jin2} J.  Jin, M. Marthaler, P.-Q. Jin, D. Golubev, and G. Sch\"{o}n,  Noise spectrum of a quantum dot–resonator 
lasing circuit, New J. Phys. {\bf 15}, 025044  (2013).

\bibitem{kar} C. Karlewski,  A. Heimes, and G. Sch\"{o}n, Lasing and transport in a multilevel double quantum dot system 
coupled to a microwave oscillator, Phys. Rev. B {\bf 93}, 045314 (2016).

\bibitem{gul2} M. J.  Gullans, J. M. Taylor, and  J. R.  Petta, Probing electron-phonon interactions in the charge-photon 
dynamics of cavity-coupled double quantum dots, Phys. Rev. B {\bf 97}, 035305 (2018).

\bibitem{st} T. M. Stace, A. C.  Doherty, and S. D. Barrett, Population Inversion of a Driven Two-Level System in a 
Structureless Bath, Phys. Rev. Lett. {\bf 95}, 106801 (2005).

\bibitem{contr} L. D.  Contreras-Pulido, C. Emary, T. Brandes, and R. Aguado,  Non-equilibrium correlations and entanglement 
in a semiconductor hybrid circuit-QED system, New J. Phys. {\bf 15}, 095008 (2013).

\bibitem{tema} T. Mihaescu, E. Cecoi, M. A. Macovei, and A. Isar, Geometric discord for a driven two-qubit system, 
Rom. Rep. Phys. {\bf 73}, 101 (2021).

\bibitem{srin} V. Srinivasa,   J. M. Taylor, and J. R.  Petta,  Cavity-mediated entanglement of parametrically driven spin qubits 
via sidebands, Phys. Rev. X Quantum {\bf 5}, 020339 (2024).

\bibitem{hell} F. Hellbach, F. Pauly, W. Belzig, and G. Rastelli, Quantum-correlated photons generated by nonlocal electron 
transport, Phys. Rev. B {\bf 105}, L241407 (2022).

\bibitem{ndqd1} C. M\"{u}ller, and Th. M. Stace, Deriving Lindblad master equations with Keldysh diagrams: 
Correlated gain and loss in higher order perturbation theory, Phys. Rev. A {\bf 95}, 013847 (2017).

\bibitem{ekm} N. A. Enaki, and M. A. Macovei, Two-photon cooperative decay in a cavity in the presence of a thermalized 
electromagnetic field, JETP {\bf  88}, 633 (1999).

\bibitem{toexp} Y. Ota, S. Iwamoto, N. Kumagai, and Y. Arakawa, Spontaneous Two-Photon Emission from a Single Quantum Dot,
Phys. Rev. Lett. {\bf 107}, 233602 (2011).

\bibitem{multp} J. Jin, M. Marthaler, and Sch\"{o}n, Electroluminescence and multiphoton effects in a resonator driven by a 
tunnel junction, Phys. Rev. B {\bf 91}, 085421 (2015). 

\bibitem{ttr1} D. F. V. James, Quantum Computation with Hot and Cold Ions: An Assessment of Proposed Schemes, 
Fort. Phys. {\bf 48}, 823 (2000).

\bibitem{ttr2} R. Tan, G.-x. Li, and Z. Ficek, Squeezed single-atom laser in a photonic crystal, 
Phys. Rev. A {\bf 78}, 023833 (2008).

\bibitem{gsag} G. S. Agarwal, {\it Quantum Statistical Theories of Spontaneous Emission and their Relation to Other Approaches } 
(Springer, Berlin, 1974).

\bibitem{kmek} M. Kiffner, M. Macovei, J. Evers, and C. H. Keitel, Vacuum induced processes in multilevel atoms, 
Prog. Opt. {\bf 55}, 85 (2010).

\bibitem{ttm} T. Quang, and H. Freedhoff, Atomic population inversion and enhancement of resonance fluorescence in a cavity,
Phys. Rev. A {\bf 47}, 2285 (1993).

\bibitem{glb} R. J. Glauber, The Quantum Theory of Optical Coherence, Phys. Rev. {\bf 130}, 2529 (1963).

\bibitem{drm} P. D. Drummond and D. F. Walls, Quantum theory of optical bistability. I. Nonlinear polarisability model, 
J. Phys. A {\bf 13}, 725 (1980).

\bibitem{mmk} M. A. Macovei, Measuring photon-photon interactions via photon detection, Phys. Rev. A {\bf 82}, 063815 (2010).

\end{thebibliography}
\end{document}